\documentclass[journal]{IEEEtran}
\usepackage[text={7.5in,10.4in}]{geometry}
\usepackage{amsmath,amssymb,amsthm}
\usepackage{graphicx}
\graphicspath{fig_MRULTRA/}
\usepackage{subfigure}
\usepackage{caption}
\usepackage{float}
\usepackage{bm}
\usepackage{diagbox}

\usepackage{algorithm}
\usepackage{algorithmicx}
\usepackage{algpseudocode}
\usepackage{url}
\usepackage{indentfirst}
\usepackage{verbatim}
\usepackage{booktabs,multirow}
\usepackage{tikz}
\usepackage{cite}
\usepackage[numbers,sort&compress]{natbib}
\usetikzlibrary{spy}
\usepackage{booktabs}
\usepackage{multirow}

\newcommand{\x}		{\mathbf{x}}	
\newcommand{\y}		{\mathbf{y}}
\newcommand{\A}		{\mathbf{A}}
\newcommand{\B}     {\mathbf{B}}

\newcommand{\W}	 {\mathbf{W}}
\renewcommand{\P}	 {\mathbf{P}}
\newcommand{\omg}	 {\mathbf{\Omega}}

\newcommand{\Z}		{\mathbf{Z}}
\newcommand{\0}	     {\mathbf{0}}

\newcommand{\D}	 {\mathbf{D}}
\newcommand{\I}	     {\mathbf{I}}

\newcommand{\s}	 {\mathbf{s}}
\newcommand{\Rsf}{\mathsf{S}}
\newcommand{\g}	 {\mathbf{g}}
\newcommand{\h}	 {\mathbf{h}}
\newcommand{\ze}	 {\bm{\zeta}}

\newcommand{\R}	     {\mathbf{R}}

\newcommand{\diag}  {\mathsf{diag}}
\newcommand{\U}		{\mathbf{U}}	
\newcommand{\V}		{\mathbf{V}}
\newcommand{\Sig}		{\mathbf{\Sigma}}

\begin{document}
%
\title{Learned Multi-layer Residual Sparsifying Transform Model for Low-dose CT Reconstruction}
%
%
%

\author{Xikai Yang, Xuehang Zheng, Yong Long$^\star$, \IEEEmembership{Member, IEEE,} Saiprasad Ravishankar, \IEEEmembership{Member, IEEE}
	\thanks{X. Yang, X. Zheng and Y. Long are with the University of Michigan–
		Shanghai Jiao Tong University Joint Institute, Shanghai Jiao Tong
		University, Shanghai 200240, China (e-mail: yangxk@sjtu.edu.cn; zhxhang@sjtu.edu.cn; yong.long@sjtu.edu.cn)}
	\thanks{S. Ravishankar is with the Department of Computational Mathematics, Science and Engineering and Department of Biomedical Engineering, Michigan State University, East Lansing, MI 48824, USA (e-mail: ravisha3@msu.edu)}
	\thanks{$^\star$Yong Long is the corresponding author.}
    \vspace{-0.2in}}

\maketitle
\begin{abstract}
Signal models based on sparse representation have received considerable attention in recent years. Compared to synthesis dictionary learning, sparsifying transform learning involves highly efficient sparse coding and operator update steps. In this work, we propose a Multi-layer Residual Sparsifying Transform (MRST) learning model wherein the transform domain residuals are jointly sparsified over layers. In particular, the transforms for the deeper layers exploit the more intricate properties of the residual maps. We investigate the application of the learned MRST model for low-dose CT reconstruction using Penalized Weighted Least Squares (PWLS) optimization. Experimental results on Mayo Clinic data show that the MRST model outperforms conventional methods such as FBP and PWLS methods based on edge-preserving (EP) regularizer and single-layer transform (ST) model, especially for maintaining some subtle details. 
\end{abstract}


%
\IEEEpeerreviewmaketitle

\section{Introduction}
Signal models exploiting sparsity have been shown to be useful in a variety of applications such as compression, restoration, denoising, reconstruction, etc. 
Natural signals can be modeled as sparse in a synthesis dictionary (i.e., represented as linear combinations of few dictionary atoms or columns) or in a sparsifying transform domain.

Transforms such as wavelets \cite{pati:93:omp} and the discrete cosine transform (DCT) are well-known to sparsify images. Methods such as synthesis dictionary learning \cite{aharon:06:ksa} and analysis dictionary learning \cite{rubinstein:13:aka} allow adapting the model to the data. These dictionary learning problems are NP-hard in general and involve computationally expensive alternating-type algorithms that limit applicability to large-scale data. In contrast, the recently proposed sparsifying transform learning approaches \cite{ravishankar:13:lst} involve exact and highly efficient updates.
 In particular, the transform model approximates the signal as sparse in a transformed domain. Adopting a multi-layer sparsifying transform model enables sparsifying an input image successively over layers \cite{ravishankar:18:lml}, creating a rich and more complete sparsity model, which forms the core of this work.

One of the most important application of such image models is for medical image reconstruction. In particular, an important problem in computed tomography (CT) is reducing the X-ray exposure to patients while maintaining good image reconstruction quality.
 A conventional method for CT reconstruction is the analytical filtered back-projection (FBP) \cite{feldkamp:84:pcb}. However, image quality degrades severely for FBP when the radiation dose is reduced. In contrast, model-based image reconstruction (MBIR) achieves comparable or better image quality \cite{elbakri:02:sir}. 

A typical MBIR method for low-dose CT (LDCT) is the penalized weighted least squares (PWLS) approach. The cost function for PWLS includes a weighted quadratic data-fidelity term and a penalty term or regularizer capturing prior information or model of the object \cite{sauer:93:alu, thibault:07:atd}. Recent works have shown promising LDCT reconstruction quality by incorporating data-driven models into the regularizer, where the models are learned from datasets of images or image patches. In particular, PWLS reconstruction with adaptive sparsifying transform-based regularization has shown promise for tomographic reconstruction \cite{pfister:14:mbi,zheng:18:pua,chun:17:esv}. Recent work has also shown that they may generalize better to new data than supervised deep learning schemes \cite{ye:19:sld}. The adaptive transform-based image reconstruction algorithms can exploit a variety of image models \cite{pfister:14:mbi,zheng:18:pua,zhou:13:atf} learned in an unsupervised manner, and involve efficient closed-form solutions for sparse coding.

In this work, we  propose a new formulation and algorithm for learning a multi-layer transform model \cite{ravishankar:18:lml}, where the transform domain residuals (the difference between transformed data and their sparse approximations) are successively sparsified over several layers. We refer to the model as a multi-layer residual sparsifying transform (MRST) model. The transforms are learned over several layers from images to jointly minimize the transform domain residuals across layers, while enforcing sparsity conditions in each layer. Importantly, the filters in deeper layers can help better exploit finer features (e.g., edges and correlations) in the residual maps. We investigate the performance of the (unsupervised) learned MRST model for LDCT reconstruction using PWLS. We propose efficient alternating minimization algorithms for both learning and reconstruction.
Experimental results on Mayo Clinic data illustrate that the learned MRST model outperforms the conventional FBP as well as PWLS methods based on the non-adaptive edge-preserving (EP) regularizer and learned single-layer transform model, especially for maintaining some subtle details. 
In the following sections, we will first discuss the proposed model and formulation, followed by our algorithms and experimental results on Mayo Clinic data.

\section{Problem Formulation}
Here, we will introduce the proposed general multi-layer learning framework and the formulation for LDCT image reconstruction. 
The MRST learning cost and constraints are shown in Problem \eqref{eq:P0}, which is an extension of simple single-layer transform learning.
\begin{equation}\label{eq:P0}
\begin{aligned}
& \quad \quad \quad \quad \min_{\{\omg_l,\Z_l\}}   \sum_{l=1}^L \bigg\{  \|\omg_l\R_l - \Z_l\|_F^2 + \eta_l^2\|\Z_l\|_0  \bigg\}
\\
& \quad \mathrm{s.t.} \quad \R_l = \omg_{l-1}\R_{l-1} - \Z_{l-1}, 2\leq l\leq L, \omg_l^T \omg_l  = \I, \forall l.
\end{aligned}
\tag{P0}
\end{equation}
Here, $\{\omg_l\in \mathbb{R}^{p \times p}\}$ and $\{\Z_l\in \mathbb{R}^{p \times N}\}$ denote the learned transforms and sparse coefficient maps for the $1 \leq l \leq L$ layers. Parameter $\eta_l$ controls the maximum allowed sparsity level (computed using the $\ell_0$ ``norm'' penalty) of $\Z_l$. The residual maps $\{\R_l\in \mathbb{R}^{p \times N}\}$ are defined in recursive form over layers, with $\R_1$ denoting the input training data. Here, we assume $\mathbf{R}_1$ to be a matrix, whose columns are (vectorized) patches drawn from image data sets. The unitary constraints for $\{\omg_l\}$ enable closed-form solutions for the sparse coefficient and transform update steps in our algorithms.
The learned MRST model can then be used to construct a data-driven regularizer in PWLS as shown in Problem \eqref{eq:P1}.
\vspace{-0.05in}
\begin{equation}\label{eq:P1}	
\min_{\x \geq \0}  \frac{1}{2}\|\y - \A \x\|^2_{\W}  + \beta \Rsf(\x),   
\tag{P1}
\end{equation}
\vspace{-0.10in}
\begin{equation*}\label{eq:Rx_DST}
\begin{aligned}
\quad \quad \quad &\Rsf(\x) \triangleq  \min_{\{\Z_l\}}  \sum_{l=1}^{L}  \bigg\{  \|\omg_l \R_l - \Z_l\|^2_F + \gamma_l^2\|\Z_l\|_0   \bigg\}  
\\ &\;   \mathrm{s.t.} \;  \R_l = \omg_{l-1}\R_{l-1} - \Z_{l-1}, 2 \leq l \leq L \, ,  \R_1^j = \P^j\x, \, \forall \, j.
\end{aligned}
\vspace{-0.05in}
\end{equation*}
In particular, we reconstruct the image $\x \in \mathbb{R}^{N_p}$ from noisy sinogram data $\y \in \mathbb{R}^{N_d}$ by solving \eqref{eq:P1}, where $\A \in  \mathbb{R}^{  N_d  \times N_p}$ is the system matrix of the CT scan and $\W = \diag \{w_i\} \in \mathbb{R}^{N_d \times N_d}$ is the diagonal weighting matrix with elements being the estimated inverse variance of $y_i$. Operator $\P^j\in \mathbb{R}^{p\times N_p}$ extracts the $j$th patch of $\x$ as $\P^j\x$. The $j$th colums of $\R_l$ and $\Z_l$ are denoted $\Z_l^j$ and $\R_l^j$. The non-negative parameters $\{\gamma_l\}$ control the sparsity of the coefficient maps in different layers, and $\beta>0$ captures the relative trade-off between the data-fidelity term and regularizer.
\vspace{-0.05in}
\section{Algorithms for Learning and Reconstruction}
\subsection{Transform Learning Algorithm}
\vspace{-0.05in}
We propose an exact block coordinate descent (BCD) algorithm for the nonconvex Problem \eqref{eq:P0} that cycles over updating each $\Z_l$ for $1 \leq l \leq L$ (sparse coding step) and each $\omg_l$ for $1 \leq l \leq L$ (transform update step). In each step, the remainder of the variables (that are not optimized) are kept fixed. The exact BCD algorithm ensures that the objective in \eqref{eq:P0} is monotone decreasing over iterations and that it converges. In particular, under the unitarity condition on the transforms, every subproblem in the block coordinate descent minimization approach can be solved exactly. We initialize the algorithm with 2D DCT for $\omg_1$ and the identity matrix for $\{\omg_l\}  (l\geq2)$ respectively. The initial $\{\Z_l\}$ are all-zero matrices.
Since the residuals are defined recursively in \eqref{eq:P0}, to simplify the algorithmic description, we first define matrices $\B_p^q (p < q)$, which can be regarded as backpropagation matrices from the $q$th to $p$th layers.
\begin{equation}\label{eq:B_pq}
\begin{aligned}
&\B_p^q=\omg_{p+1}^T\Z_{p+1}+\omg_{p+1}^T\omg_{p+2}^T\Z_{p+2}+...+\omg_{p+1}^T\omg_{p+2}^T...\omg_q^T\Z_q
\\
&=\sum_{k=p+1}^q \bigg( \prod_{s=p+1}^k \omg_s^T \bigg)\Z_k
\end{aligned} 
\end{equation}
\subsubsection{Sparse Coding Step for $\Z_l$}
Here, we solve (P0) for $\Z_l$ with all other variables fixed. The corresponding subproblem is as in \eqref{eq:sub_pro_Z}.
\begin{equation}\label{eq:sub_pro_Z}
\vspace{-0.05in}
\min_{\Z_l}   \sum_{i=l}^L \bigg\{  \|\omg_i\R_i - \Z_i\|_F^2 \bigg\}  + \eta_l^2\|\Z_l\|_0
\end{equation}	
Although the cost function is nonconvex and the residual maps also depend on $\Z_l$ (as in \eqref{eq:P0}), we can exploit the unitarity of the transforms to simplify and rewrite Problem \eqref{eq:sub_pro_Z} as $\min_{\Z_l}(L-l+1)\times\|\Z_l - ( \omg_l\R_l - \frac{1}{L-l+1}\sum_{i=l+1}^L \B_l^i )\|_F^2 + \eta_{l}^{2}\|\Z_l\|_0$. This problem has a similar form as the single-transform sparse coding problem \cite{ravishankar:13:lst}, and the optimal solution $\hat{\Z}_l$ is obtained as in \eqref{eq:Z_l}, where $H_{\eta}(\cdot)$ denotes the \textit{hard-thresholding} operator that sets elements with magnitude less than the threshold $\eta$ to zero.
\begin{equation}\label{eq:Z_l}
\begin{aligned}
\hat{\Z}_l=
&\begin{cases}
H_{\eta_l/\sqrt{L-l+1}} \bigg( \omg_l\R_l - \frac{1}{L-l+1}\sum_{i=l+1}^L \B_l^i \bigg),&  1\leq l \leq L-1,\\
H_{\eta_L} (\omg_L\R_L),& l=L.
\end{cases}
\end{aligned}
\end{equation}
%
\subsubsection{Transform Update Step for $\omg_l$}
Here, we fix $\{\Z_l\}$ and all $\omg_l$ (except the target $\omg_l$ in \eqref{eq:P0}) and solve the following subproblem:
\vspace{-0.05in}
\begin{equation}\label{eq:sub_pro_mOmega}
\min_{\omg_l}   \sum_{i=l}^L \bigg\{ \|\omg_i\R_i - \Z_i\|_F^2 \bigg\}
\quad \mathrm{s.t.} \quad  \omg_l^T \omg_l  = \I. 
\end{equation}	

Incorporating the recursive dependence of the residual maps on the transforms (as in \eqref{eq:P0}) along with the unitarity of the transforms simplifies the solution to \eqref{eq:sub_pro_mOmega}. First, denoting the full singular value decomposition (SVD) of the matrix $\mathbf{G}_l$ below by
 $\U_l\Sig_l\V_l^T$, the optimal solution to \eqref{eq:sub_pro_mOmega} is $\hat{\omg_l}=\V_l\U_l^T$.
\vspace{-0.05in}
\begin{equation}\label{eq:mOmega_l}
\mathbf{G}_l =
\begin{cases}
\R_l\bigg( \Z_l + \frac{1}{L-l+1} \sum_{i=l+1}^L \B_l^i \bigg)^T, & 1\leq l \leq L-1,\\
\R_L\Z_L^T, & l=L.
\end{cases}
\vspace{-0.05in}
\end{equation}

%
%
%

\subsection{Image Reconstruction Algorithm}
The proposed PWLS-MRST algorithm for low-dose CT image reconstruction exploits the learned MRST model. We reconstruct the image by solving the PWLS problem \eqref{eq:P1}. We propose a block coordinate descent (BCD) algorithm for \eqref{eq:P1} that cycles over updating the image $\x$ and each of the sparse coefficient maps $\Z_l$ for $1 \leq l \leq L$.
\subsubsection{Image Update Step for $\x$}
First, with the sparse coefficient maps $\{\Z_l\}$ fixed, we optimize for $\x$ in \eqref{eq:P1} by optimizing the following subproblem:
\begin{equation} \label{eq:sub_P1}
\min_{\x \geq \0} \frac{1}{2} \|\y - \A\x \|^2_{\W} + \beta \Rsf_2(\x),   
\end{equation}
where $\Rsf_2(\x) \triangleq   \sum_{l=1}^{L}  \bigg\{ \|\omg_l\R_l - \Z_l\|_F^2 \bigg\} $, with $\R_l=\omg_{l-1}\R_{l-1} - \Z_{l-1}$, $2 \leq l \leq L$, and $\R_1^j = \P^j \x$.	
We use the \textit{efficient} relaxed linearized augmented Lagrangian method with ordered-subsets (relaxed OS-LALM) algorithm \cite{nien:16:rla} to obtain the solution to \eqref{eq:sub_P1}. The algorithmic details are shown in \textbf{Algorithm 1}. In each iteration of the OS-LALM method, we update the image $M$ times corresponding to $M$ ordered subsets. The matrices $\A_m$ and $\W_m$, and the vector $\y_m$ are sub-matrices of $\A$ and $\W$, and sub-vector of $\y$ for the $m$th subset respectively. Matrix $\D_{\A}$ represents a diagonal majorizing matrix of $\A^T\W\A$. We precompute the Hessian matrix of  $\Rsf_2(\x)$ as $\D_{\Rsf_2}$ in \eqref{eq:D_R} to accelerate the algorithm, and the gradient of $\Rsf_2(\x)$ is as in \eqref{eq:grad_R}.
\vspace{-0.05in}
\begin{equation}\label{eq:grad_R}		
\nabla \Rsf_2(\x)= 2 \beta \sum_{j=1}^N (\P^T)^j \bigg\{ L\P^j\x - \sum_{k=1}^L (\B_0^k)^j \bigg\}
\end{equation}	
\vspace{-0.05in}
\begin{equation}\label{eq:D_R}		
\D_{\Rsf_2}  \triangleq  \nabla^2 \Rsf_2(\x) = 2L \beta \sum_{j=1}^{N}  (\P^T)^j\P^j \\
\vspace{-0.05in}
\end{equation}
\subsubsection{Sparse Coding Step for $\Z_l$}
Similar to the sparse coding step during transform learning, the solution of \eqref{eq:P1} with respect to each sparse coefficient map $\Z_l$ is shown in \eqref{eq:recon_formula_z_l}, and is the solution of \eqref{eq:recon_pro_z_l}.
\vspace{-0.10in}
\begin{equation}\label{eq:recon_pro_z_l}
\begin{aligned}
\min_{\Z_l} \sum_{i=l}^L \bigg\{  \|\omg_i\R_i - \Z_i\|_F^2 \bigg\}  + \gamma_l^2\|\Z_l\|_0  
\\
\quad \mathrm{s.t.} \quad \R_i=\omg_{i-1}\R_{i-1} - \Z_{i-1}, l \leq i \leq L
\end{aligned}
\end{equation}		
\vspace{-0.05in}
\begin{equation}\label{eq:recon_formula_z_l}
\hat{\Z}_l=H_{\eta_l/\sqrt{L-l+1}} \bigg\{ \omg_l\R_l - \frac{1}{L-l+1}\sum_{i=l+1}^L \B_l^i \bigg\}
\end{equation}

\begin{algorithm}[!h]  
	\caption{Image Reconstruction Algorithm}\label{alg: mrst_recon}
	\begin{algorithmic}[0]
		\State \textbf{Input:}
		initial image $\tilde{\x}^{(0)}$, all-zero initial $\{\tilde{\Z_l}^{(0)}\}$, pre-learned $\{\omg_l\}$, thresholds $\{\gamma_l\}$, $\alpha = 1.999$,
		$\D_{\A} $, $\D_{\Rsf_2}$, number of outer iterations $T_{O}$, number of inner iterations $N$, and number of subsets $M$.											
		\State \textbf{Output:}  reconstructed image $\tilde{\x}^{(T)}$.
		\For {$t =0,1,2,\cdots,{T_{O}-1}$}		
		
		\State \textbf{1) Image Update}: $\{\tilde{\Z_l}^{(t)}\}$ fixed,	
		
		\textbf{Initialization:} $\rho=1$, $\x^{(0)} = \tilde{\x}^{(t)}$, $\g^{(0)} = \ze^{(0)}  = M\A_M^{T}\W_M(\A_M\x^{(0)}-\y_M) $ and $\h^{(0)} = \D_\A \x^{(0)} - \ze^{(0)}$.

		\For {$n =0,1,2,\cdots,N-1$}	
		\For {$m =0,1,2,\cdots,M-1$}  $r = nM +m$			
		\begin{equation*}
		\left\{			
		\begin{aligned}
		\s^{(r+1)} &= \rho(\D_\A \x^{(r)} -\h^{(r)}) + (1-\rho)\g^{(r)} \\
		\x^{(r+1)} &= [\x^{(r)} - (\rho\D_\A+\D_{\Rsf_2})^{-1}(\s^{(r+1)} +\nabla \Rsf_2(\x^{(r)}))]_+ \\
		\ze^{(r+1)} & \triangleq  M \A^T_m\W_m(\A_m\x^{(r+1)}-\y_m)   \\
		\g^{(r+1)} &= \frac{\rho}{\rho+1}(\alpha \ze^{(r+1)} + (1-\alpha)\g^{(r)}) +  \frac{1}{\rho+1}\g^{(r)}\\
		\h^{(r+1)}  &= \alpha(\D_{\A} \x^{(r+1)} -\ze^{(r+1)}) + (1-\alpha)\h^{(r)} 
		\end{aligned}
		\right.
		\end{equation*}  
		\State decreasing $\rho$. 
		\EndFor			
		\EndFor	
		\State   $\tilde{\x}^{(t+1)} = \x^{(NM)}$. 
		\State \textbf{2) Sparse Coding}: with $\tilde{\x}^{(t+1)}$ fixed, for each $1\leq l \leq L$, update $\tilde{\Z_l}^{(t+1)}$ by \eqref{eq:recon_formula_z_l}.
		\EndFor	
	\end{algorithmic}
\end{algorithm}
\vspace{-0.20in}
\section{Experimental Results}
In this section, we evaluate the image reconstruction quality for the proposed PWLS-MRST algorithm and compare it with conventional \textbf{FBP} and the \textbf{PWLS-EP} (Edge Preserving) method \cite{cho:15:rdf}.
\vspace{-0.10in}
\subsection{Transform Learning}
\begin{figure}[!h]
	\centering    
	\subfigure[ST] {
		\label{fig:tran_ST}     
		\includegraphics[width=0.15\textwidth]{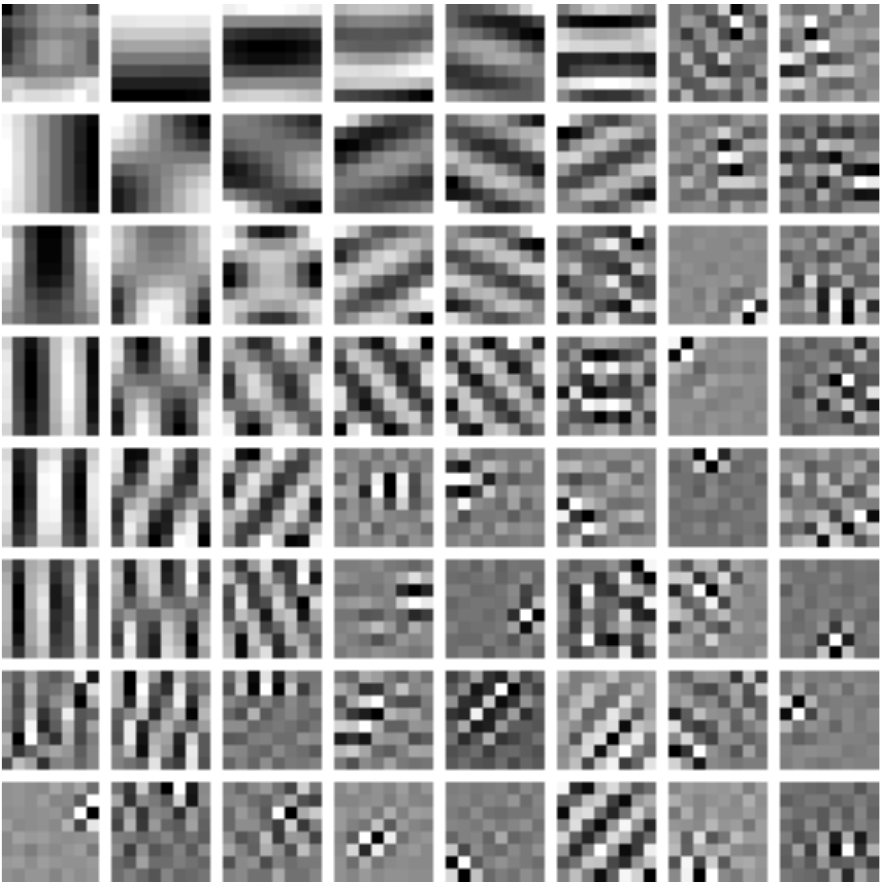} 
	}     
	\subfigure[MRST2] { 
		\label{fig:tran_MRST2}     
		\includegraphics[width=0.15\textwidth]{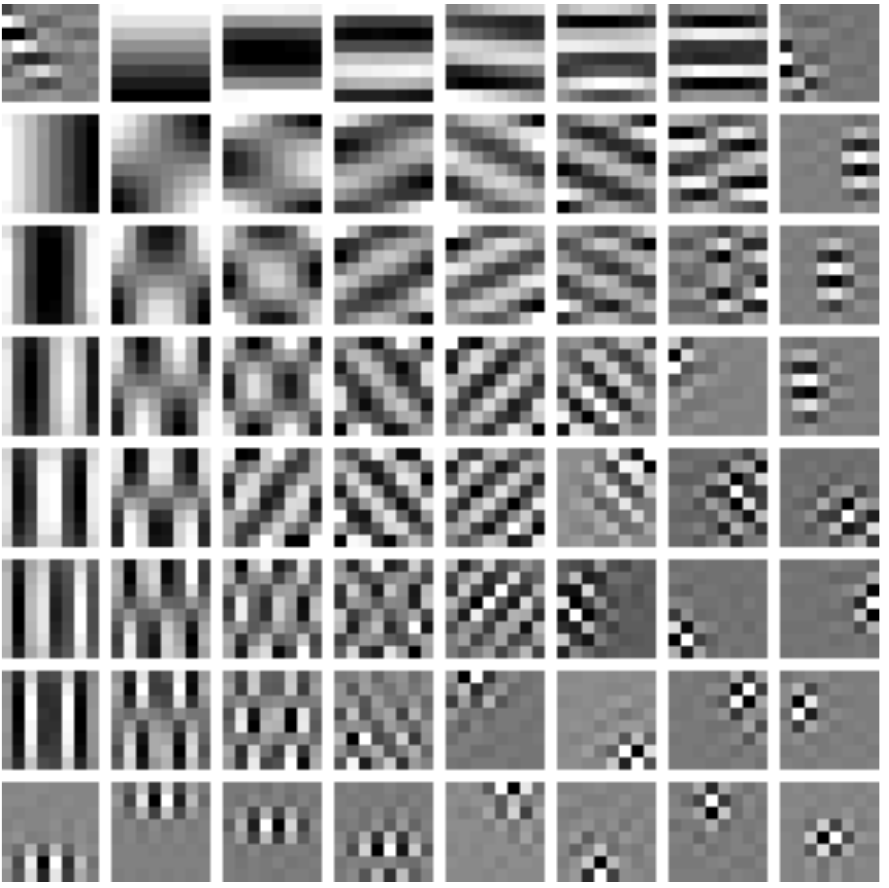}
		\includegraphics[width=0.15\textwidth]{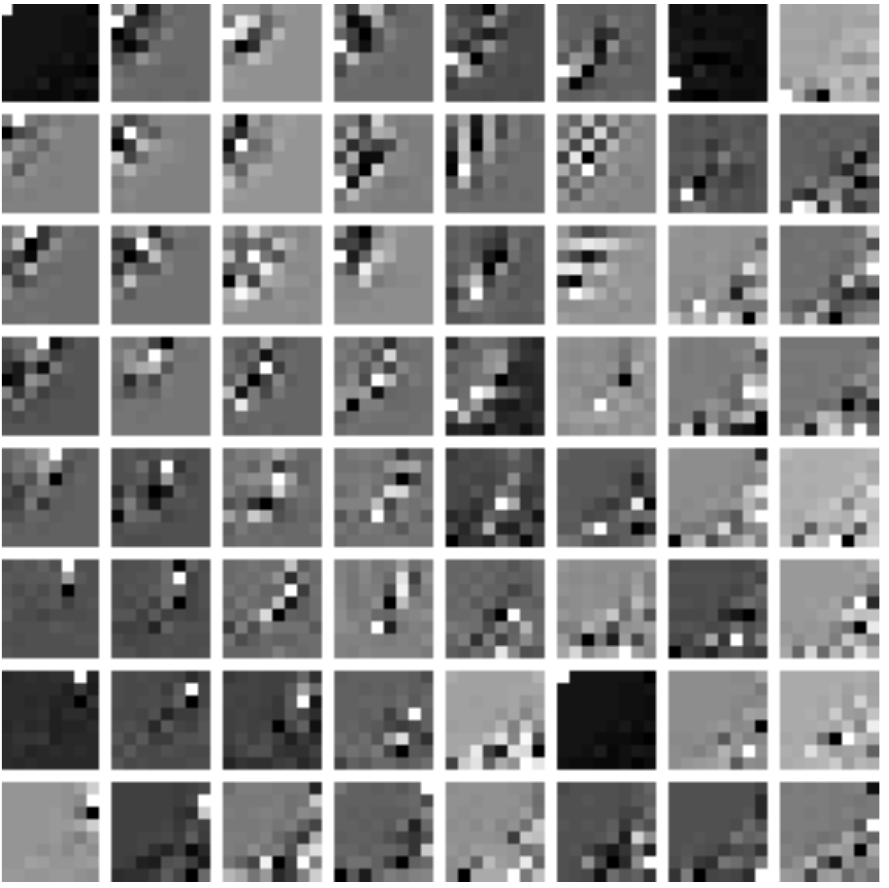}
	}    
	\subfigure[MRST3] { 
		\label{fig:tran_MRST3}     
		\includegraphics[width=0.15\textwidth]{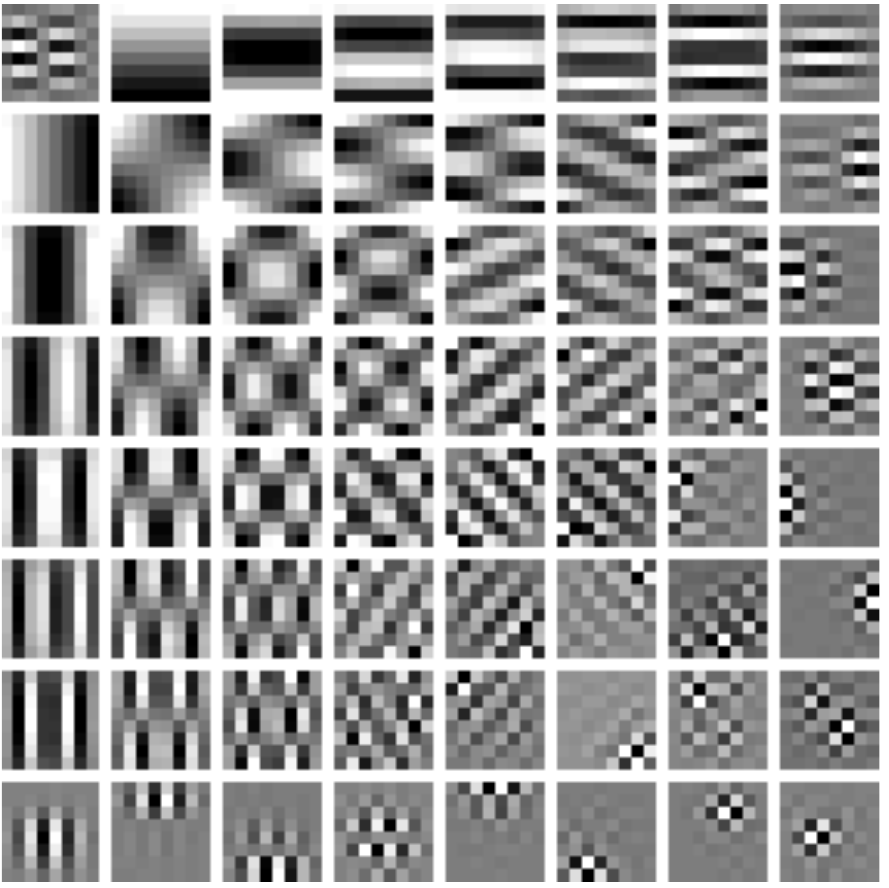}  
		\includegraphics[width=0.15\textwidth]{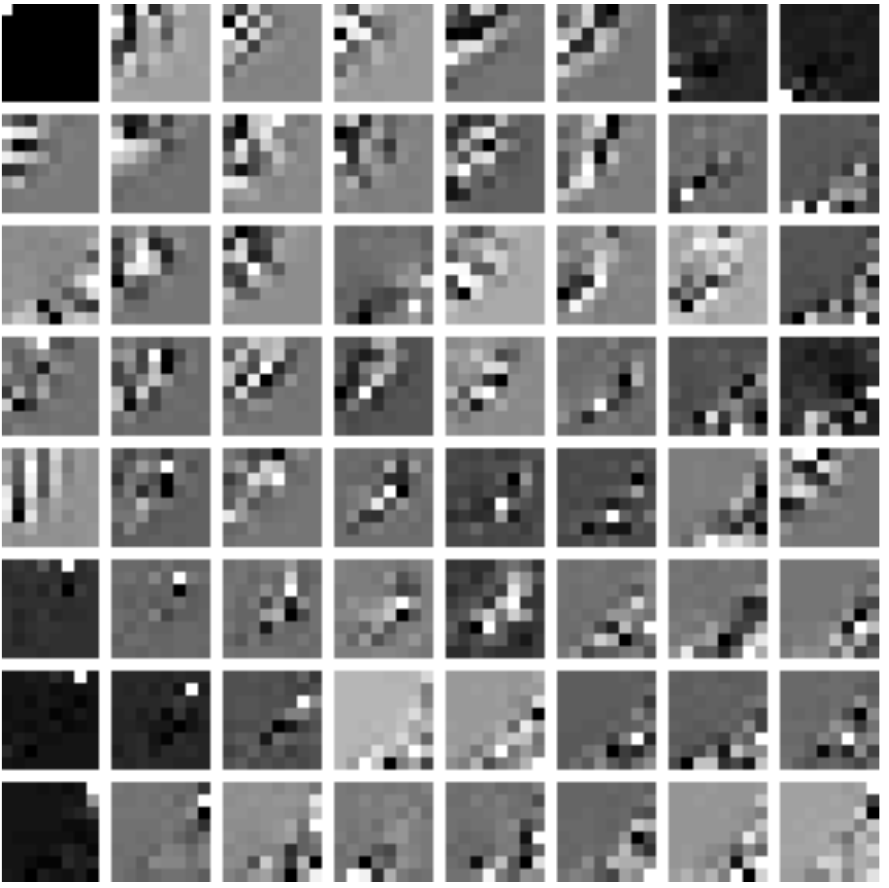}
		\includegraphics[width=0.15\textwidth]{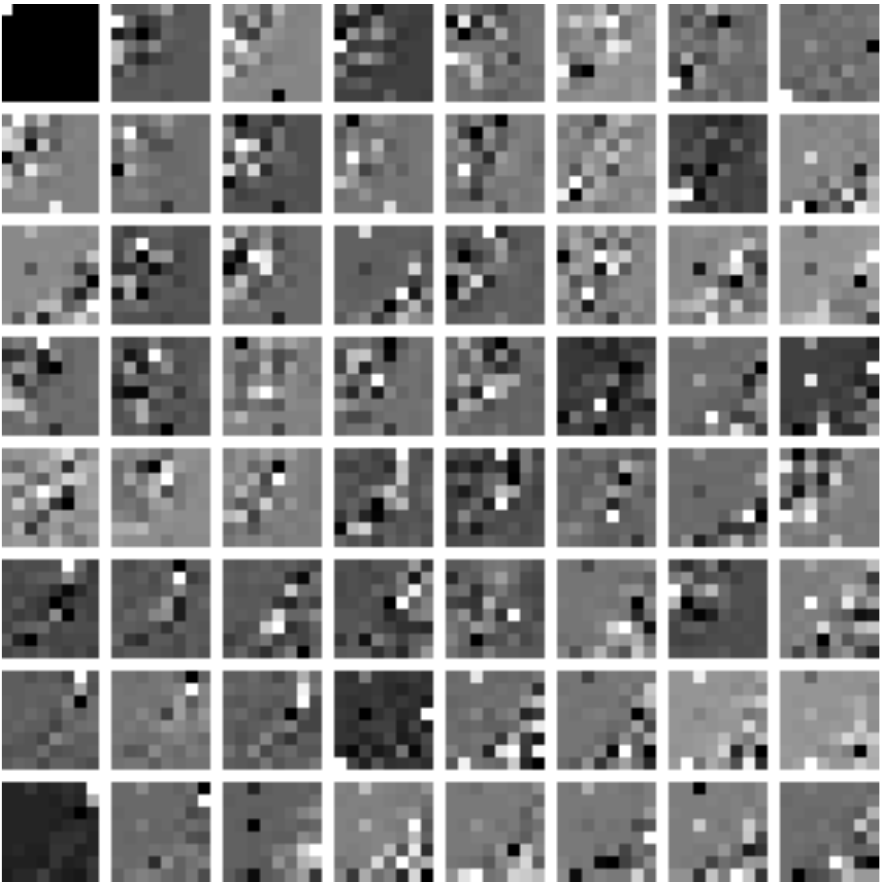}     
	}   
	\vspace{-0.10in} 
	\caption{Pre-learned transforms for ST, MRST2, and MRST3 (each transform matrix row is shown as a square patch).}     
	\label{fig:lear_tran} 
	\vspace{-0.25in}    
\end{figure}

To better understand the potential of the MRST model, we vary the number of layers and pre-learn transforms for ST, MRST2, MRST3, MRST5, and MRST7, which possess $1$, $2$, $3$, $5$, and $7$ layers, respectively. We used 7 slices of the Mayo Clinic data to train the models. For each model, we run 1000 to 2000 iterations of the learning algorithm to ensure convergence. Fig.~\ref{fig:lear_tran} shows some of the learned transforms, with each transform matrix row displayed as a square patch for simplicity. The single layer transform displays edge-like and directional structures that sparsify the image. However, with more layers, finer level features are learned to sparsify transform domain residuals in deeper layers. Nonetheless, transforms in deep layers could be more easily contaminated with noise in the training data, since the main image features are successively filtered out over layers.
\vspace{-0.10in}

\vspace{-0.10in}
\subsection{Image Reconstruction}

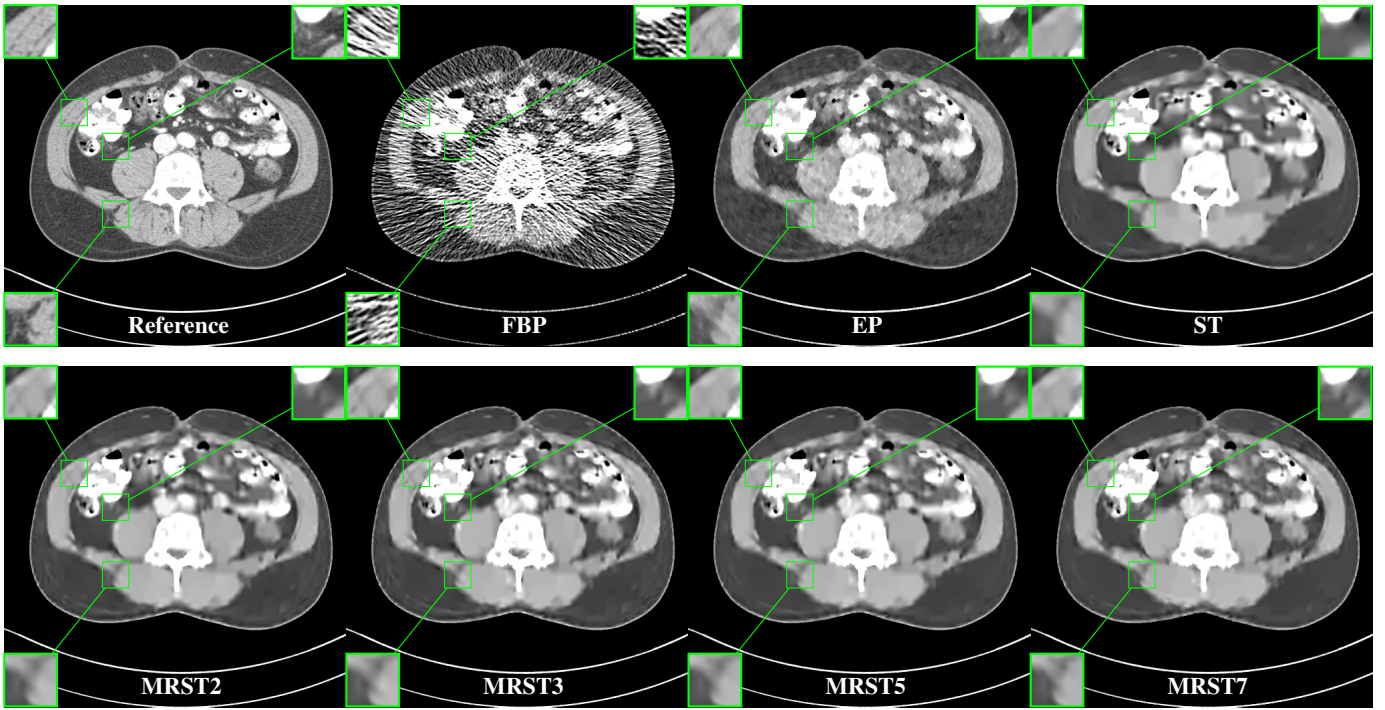
\begin{figure*}[!h]
	\vspace{-0.2in}
	\centering  
	\begin{tikzpicture}
	[spy using outlines={rectangle,green,magnification=2,size=7mm, connect spies}]
	\node {\includegraphics[width=0.24\textwidth]{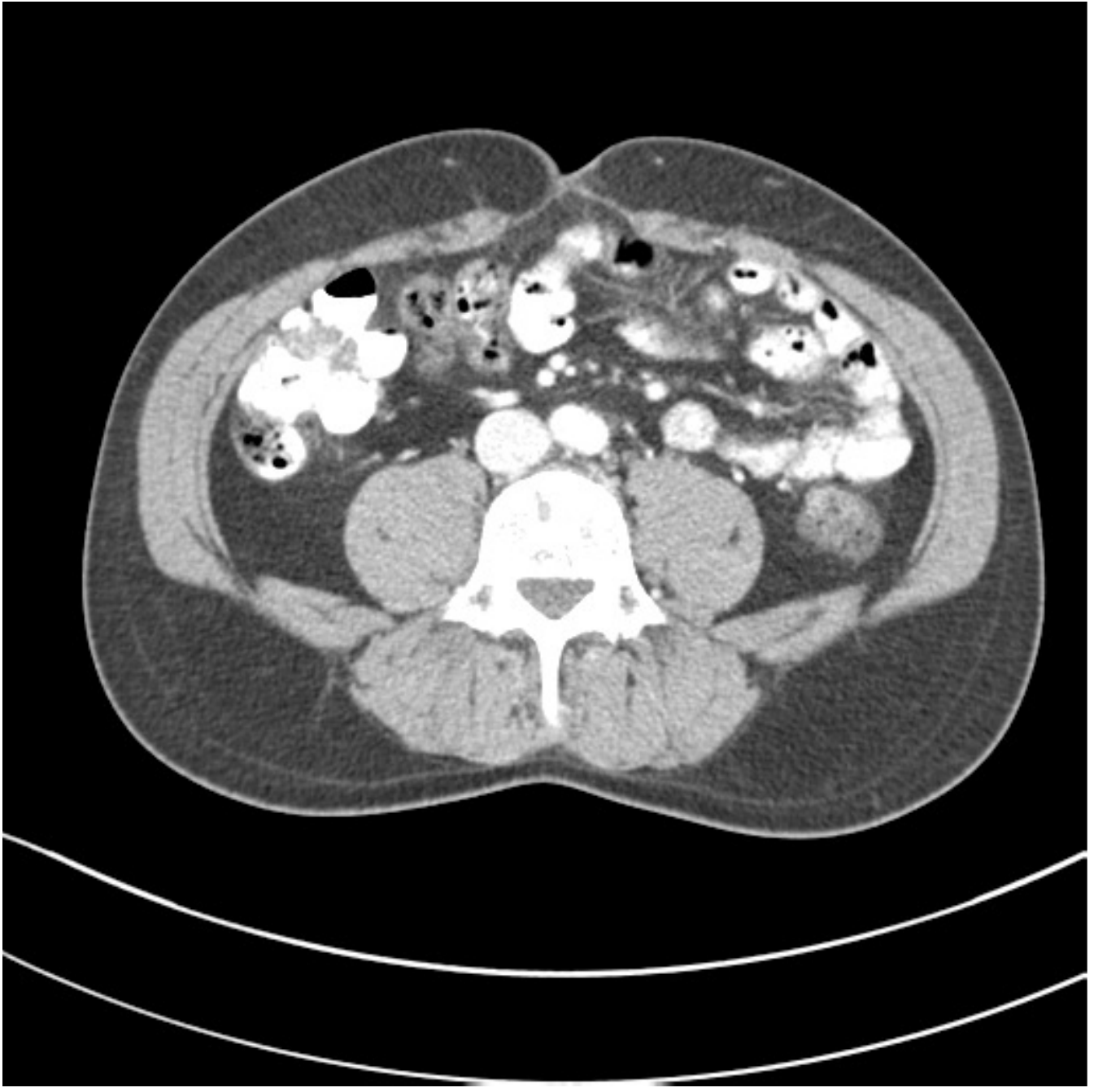}	};
	\spy on (-1.35,0.85) in node [left] at (-1.58,1.93);
	\spy on (-0.80,-0.5) in node [left] at (-1.58,-1.9);
	\spy [width=7mm, height =7mm] on (-0.80,0.40) in node [left] at (2.25,1.92);
	\end{tikzpicture}
	\put(-90,9){ \color{white}{\bf \small{Reference}}} 
	\hspace{-0.15in}
	\begin{tikzpicture}
	[spy using outlines={rectangle,green,magnification=2,size=7mm, connect spies}]
	\node {\includegraphics[width=0.24\textwidth]{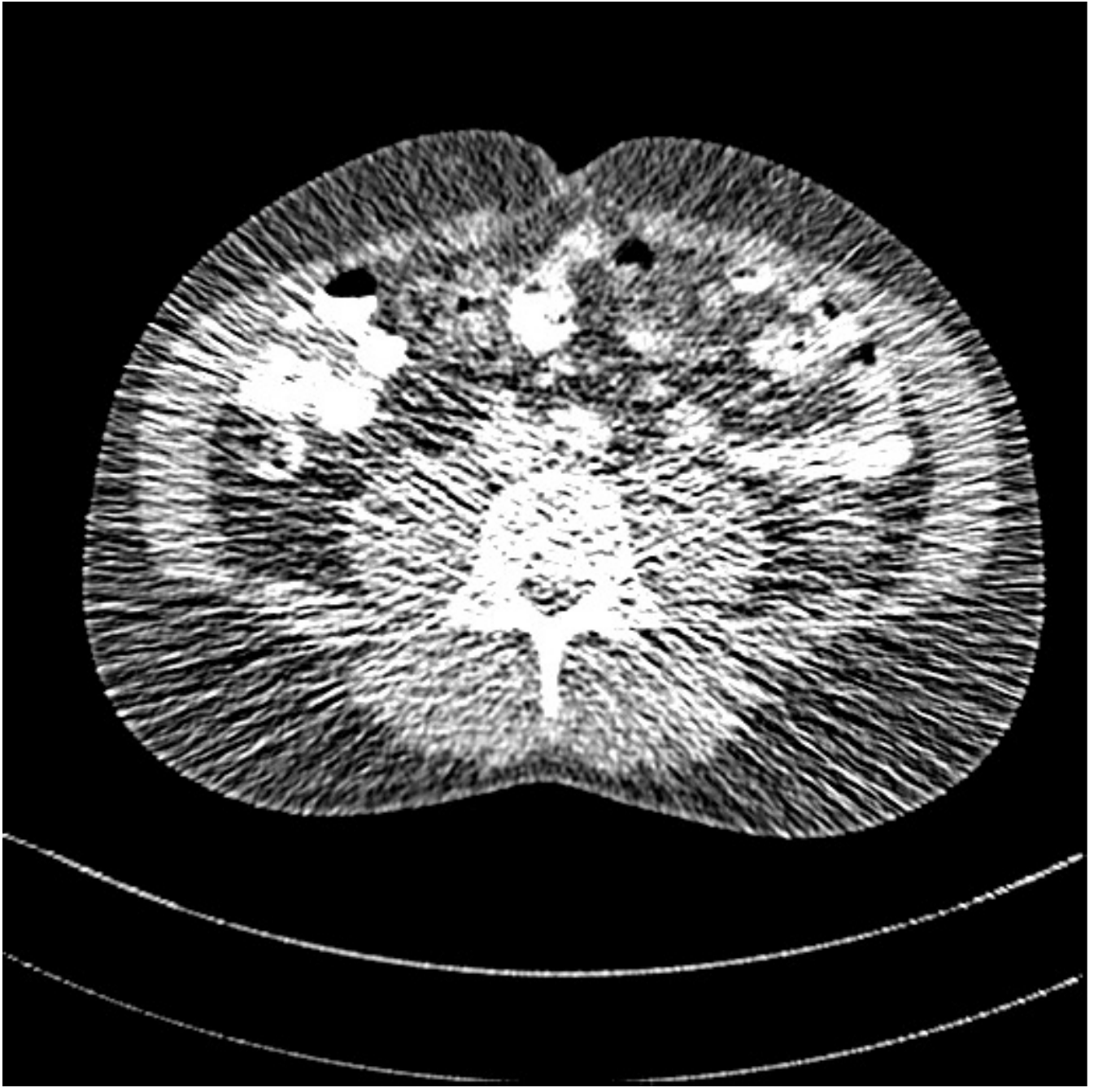}	};
	\spy on (-1.35,0.85) in node [left] at (-1.58,1.93);
	\spy on (-0.80,-0.5) in node [left] at (-1.58,-1.9);
	\spy [width=7mm, height =7mm] on (-0.80,0.40) in node [left] at (2.25,1.92);
	\end{tikzpicture}
	\put(-78,9){ \color{white}{\bf \small{FBP}}}
	\hspace{-0.15in}
	\begin{tikzpicture}
	[spy using outlines={rectangle,green,magnification=2,size=7mm, connect spies}]
	\node {\includegraphics[width=0.24\textwidth]{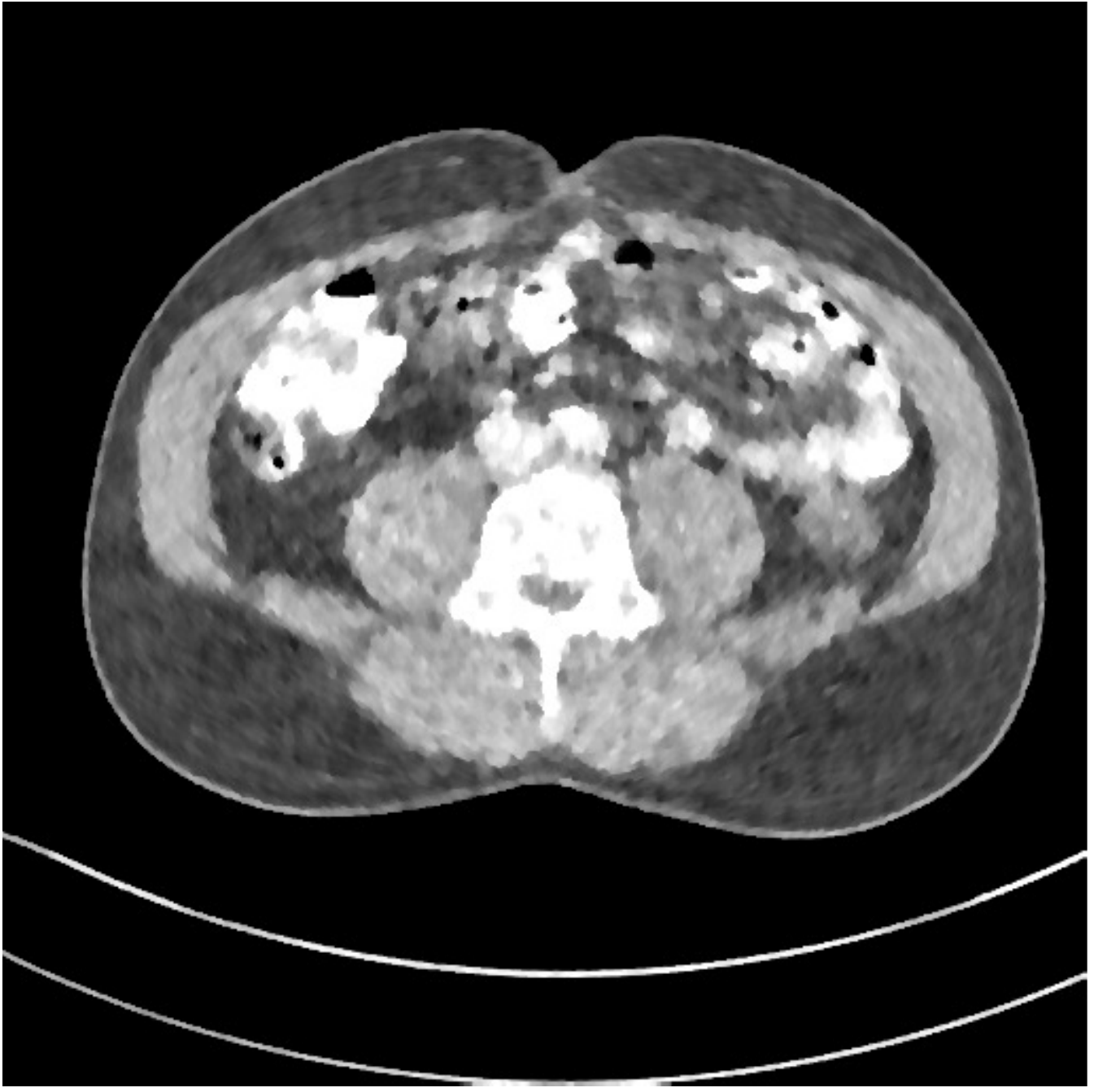}	};
	\spy on (-1.35,0.85) in node [left] at (-1.58,1.93);
	\spy on (-0.80,-0.5) in node [left] at (-1.58,-1.9);
	\spy [width=7mm, height =7mm] on (-0.80,0.40) in node [left] at (2.25,1.92);
	\end{tikzpicture}	
	\put(-75,9){ \color{white}{\bf \small{EP}}}
	\hspace{-0.15in}
	\begin{tikzpicture}
	[spy using outlines={rectangle,green,magnification=2,size=7mm, connect spies}]
	\node {\includegraphics[width=0.24\textwidth]{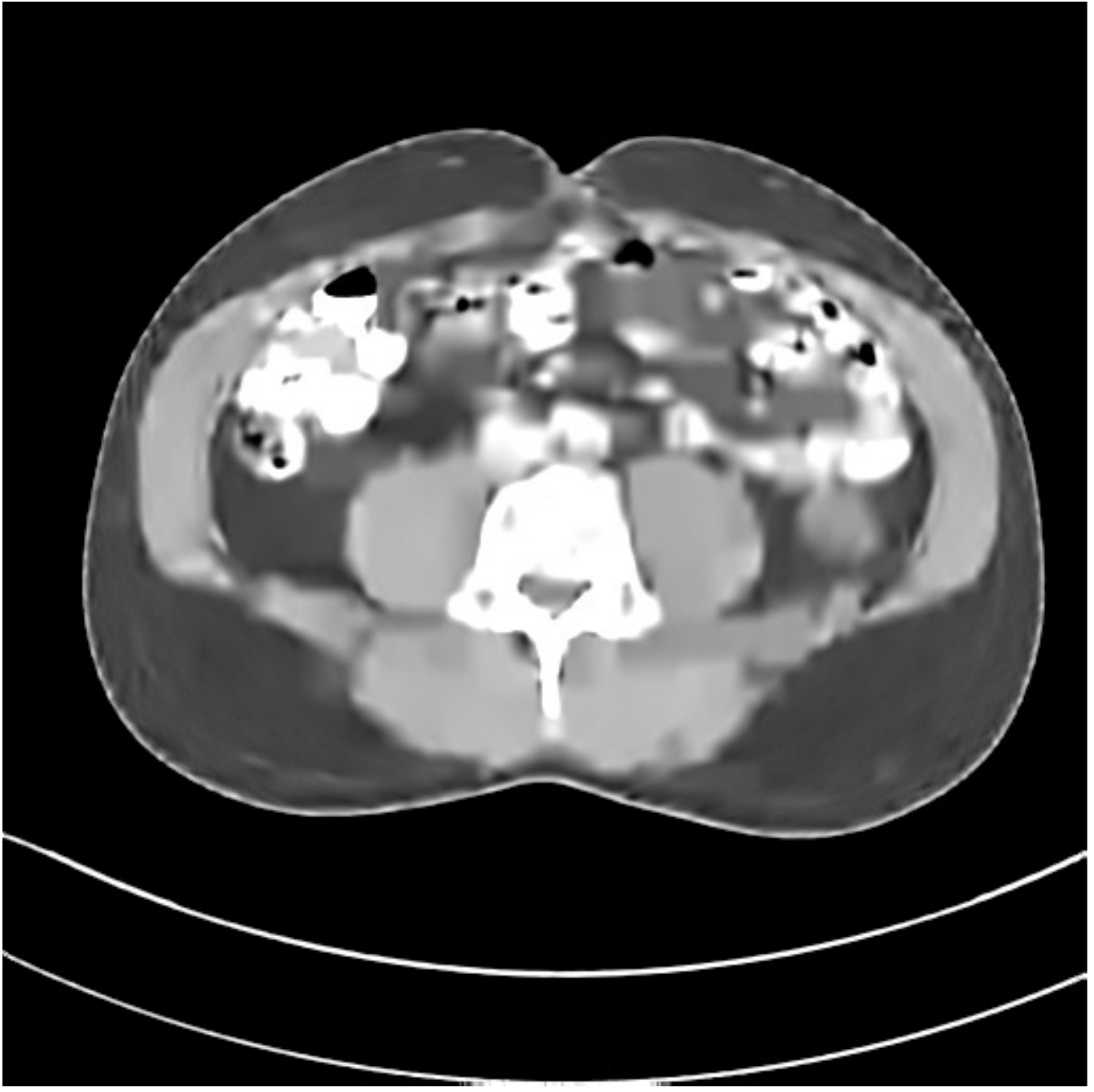}	};
	\spy on (-1.35,0.85) in node [left] at (-1.58,1.93);
	\spy on (-0.80,-0.5) in node [left] at (-1.58,-1.9);
	\spy [width=7mm, height =7mm] on (-0.80,0.40) in node [left] at (2.25,1.92);
	\end{tikzpicture}
	\put(-75,9){ \color{white}{\bf \small{{ST}}}}
	\\
	\begin{tikzpicture}
	[spy using outlines={rectangle,green,magnification=2,size=7mm, connect spies}]
	\node {\includegraphics[width=0.24\textwidth]{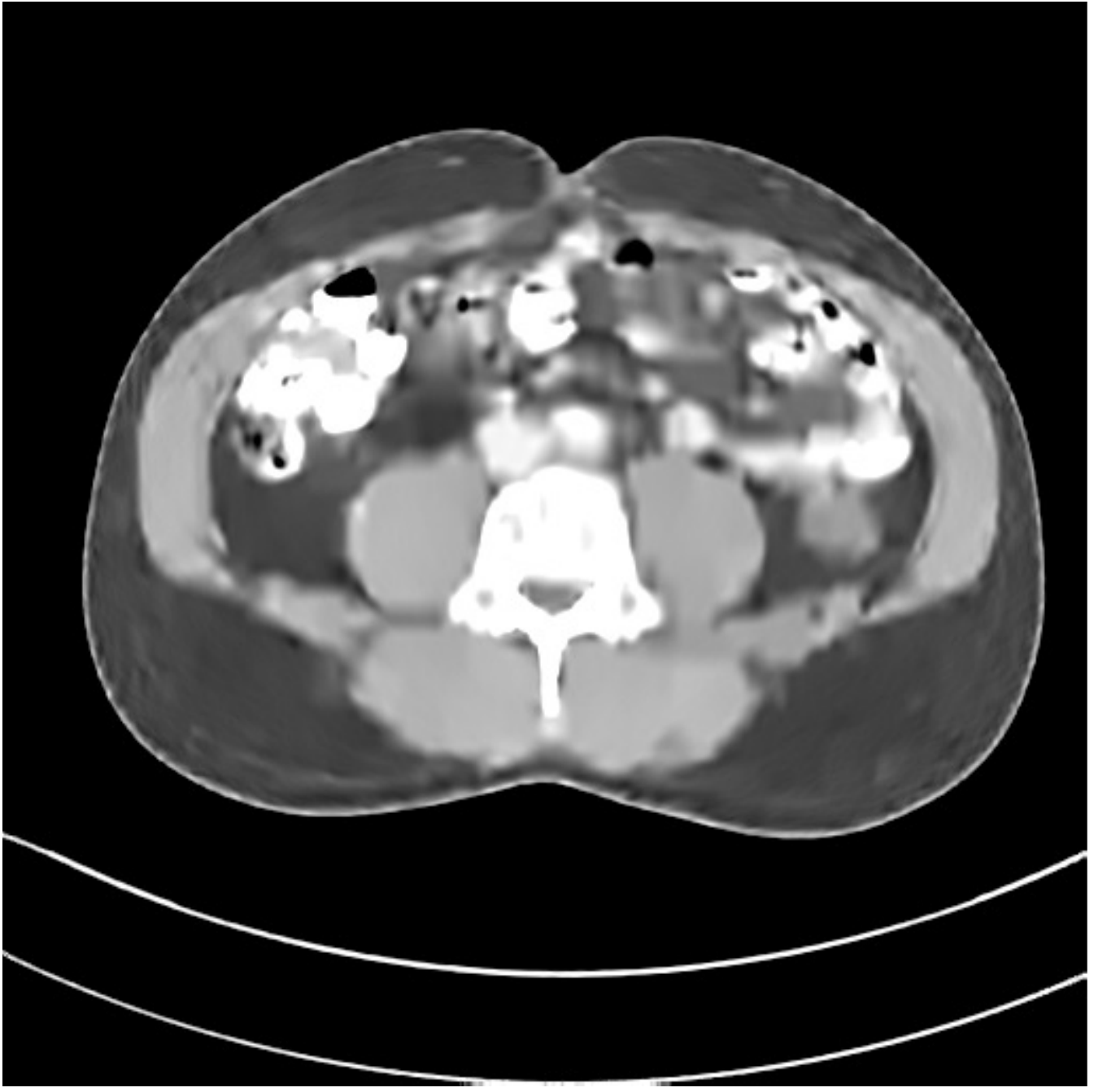}	};
	\spy on (-1.35,0.85) in node [left] at (-1.58,1.93);
	\spy on (-0.80,-0.5) in node [left] at (-1.58,-1.9);
	\spy [width=7mm, height =7mm] on (-0.80,0.40) in node [left] at (2.25,1.92); 
	\end{tikzpicture}
	\put(-85,9){ \color{white}{\bf \small{MRST2}}}
	\hspace{-0.15in}
	\begin{tikzpicture}
	[spy using outlines={rectangle,green,magnification=2,size=7mm, connect spies}]
	\node {\includegraphics[width=0.24\textwidth]{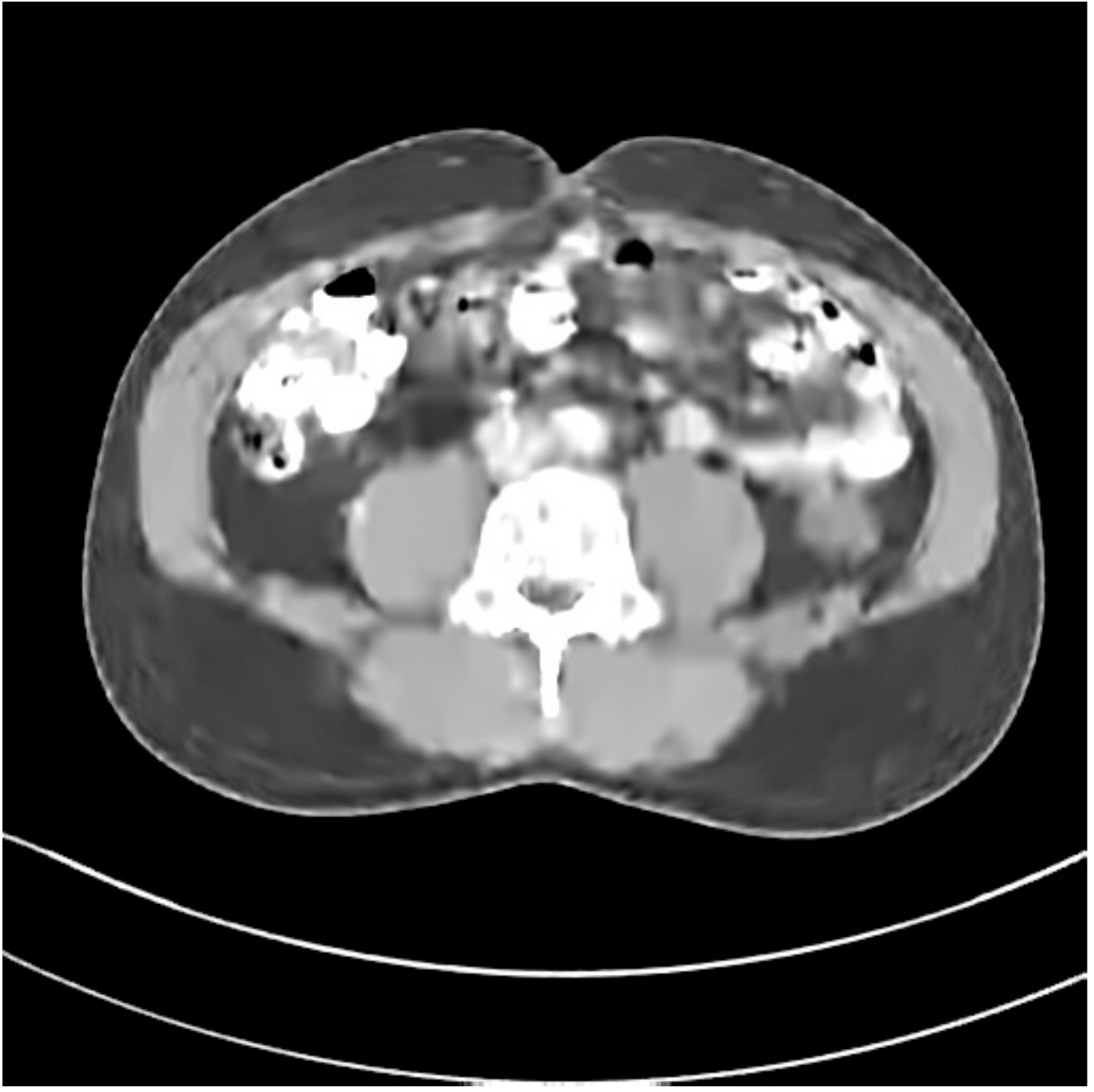}	};
	\spy on (-1.35,0.85) in node [left] at (-1.58,1.93);
	\spy on (-0.80,-0.5) in node [left] at (-1.58,-1.9);
	\spy [width=7mm, height =7mm] on (-0.80,0.40) in node [left] at (2.25,1.92);
	\end{tikzpicture}
	\put(-85,9){ \color{white}{\bf \small{MRST3}}}
	\hspace{-0.15in}
	\begin{tikzpicture}
	[spy using outlines={rectangle,green,magnification=2,size=7mm, connect spies}]
	\node {\includegraphics[width=0.24\textwidth]{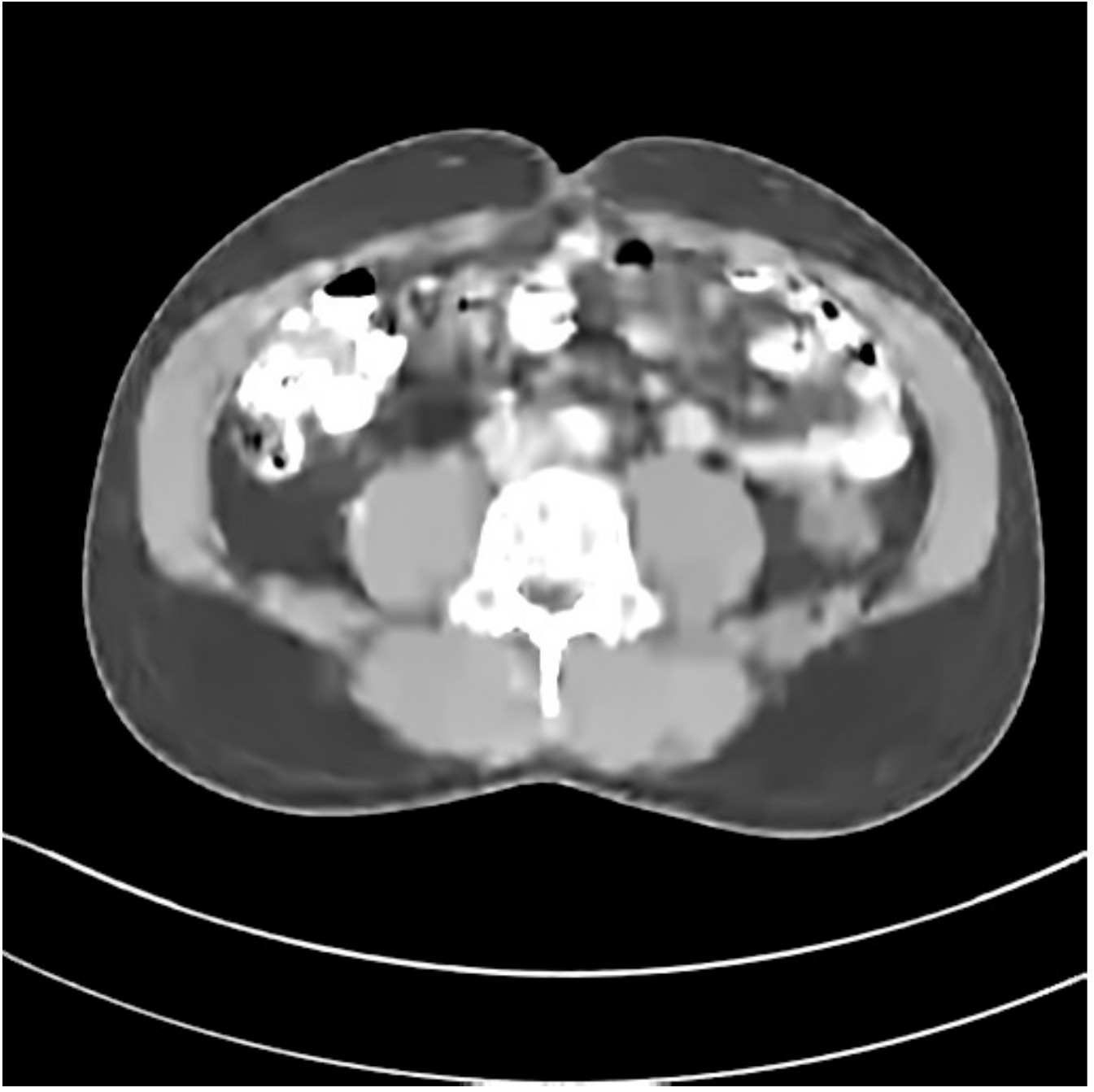}	};
	\spy on (-1.35,0.85) in node [left] at (-1.58,1.93);
	\spy on (-0.80,-0.5) in node [left] at (-1.58,-1.9);
	\spy [width=7mm, height =7mm] on (-0.80,0.40) in node [left] at (2.25,1.92);
	\end{tikzpicture}
	\put(-85,9){ \color{white}{\bf \small{MRST5}}}
	\hspace{-0.15in}
	\begin{tikzpicture}
	[spy using outlines={rectangle,green,magnification=2,size=7mm, connect spies}]
	\node {\includegraphics[width=0.24\textwidth]{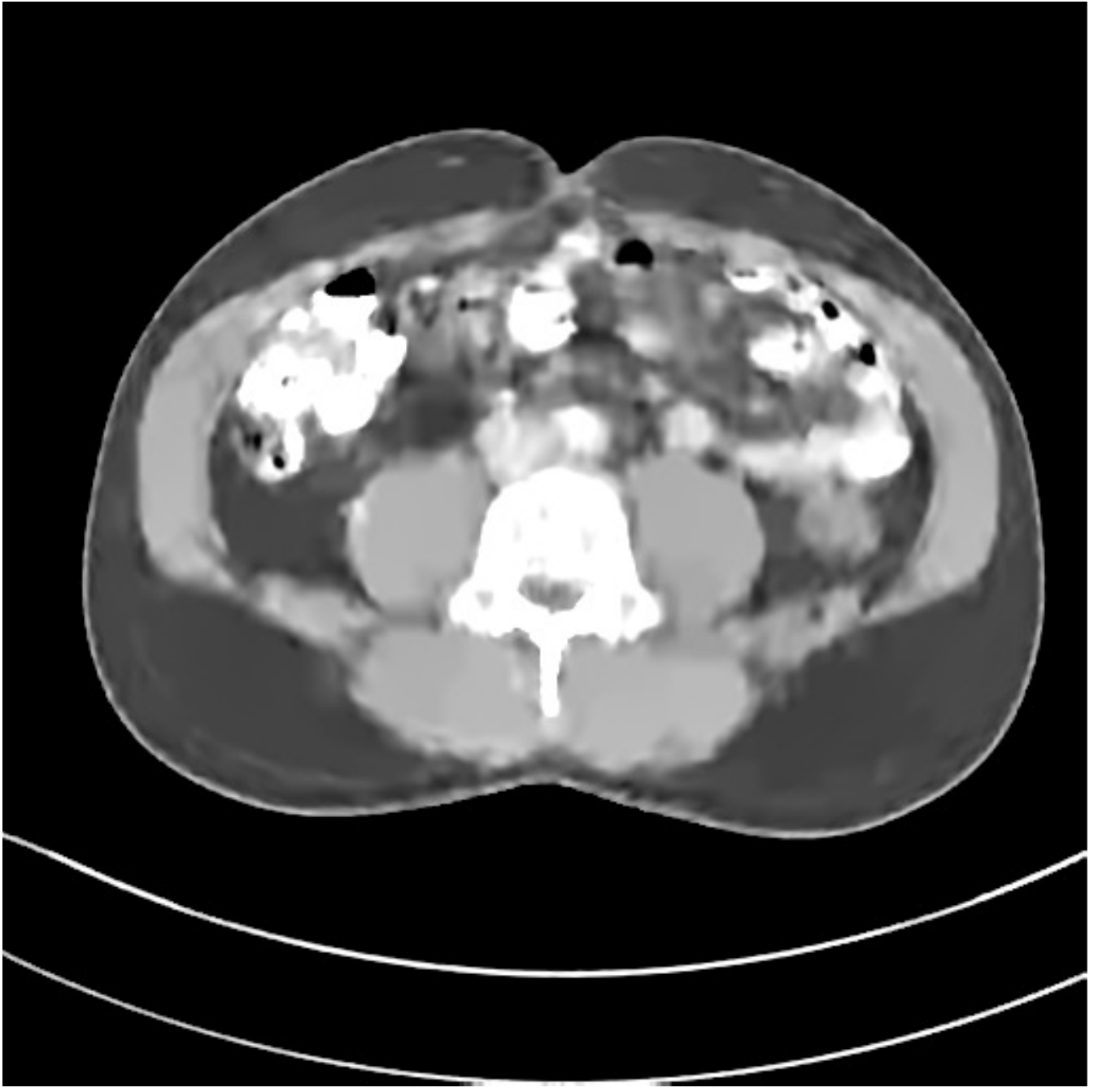}	};
	\spy on (-1.35,0.85) in node [left] at (-1.58,1.93);
	\spy on (-0.80,-0.5) in node [left] at (-1.58,-1.9);
	\spy [width=7mm, height =7mm] on (-0.80,0.40) in node [left] at (2.25,1.92);
	\end{tikzpicture}
	\put(-85,9){ \color{white}{\bf \small{MRST7}}}
	\caption{Comparison of reconstructions of L067-slice100 with FBP, PWLS-EP, PWLS-ST, PWLS-MRST2, PWLS-MRST3, PWLS-MRST5, and PWLS-MRST7, respectively at incident photon intensities $I_0=10^{4}$. The display window is [800, 1200] HU.}
	\label{fig:recon_mayo}
	\vspace{-0.15in}
\end{figure*}

Here, we simulated low-dose CT measurements from full-dose CT images in Mayo Clinic dataset with GE 2D LightSpeed fan-beam geometry corresponding to a monoenergetic source with $10^4$ incident photons per ray and no scatter.
We ran 100 iterations of the PWLS-EP algorithm with FBP reconstructions as initializations. We used the relaxed OS-LALM algorithm with $4$ ordered subsets and regularization parameter $\beta = 2^{15.5}$. For the MRST model, we used the OS-LALM algorithm for the image update step with $2$ inner iterations and $4$ subsets. We used $T_O = 1000$ and $T_O = 1500$ outer iterations for (ST, MRST2) and (MRST3, MRST5, MRST7), respectively. We set the regularization parameters (after tuning over ranges of values) as $(\beta$, $\gamma)$ $=$ $(7\times10^4$, $20)$ for ST, $(\beta$, $\gamma_1$, $\gamma_2)$ $=$ $(3\times10^4$, $30$, $12)$ for MRST2, $(\beta$, $\gamma_1$, $\gamma_2$, $\gamma_3)$ $=$ $(3\times10^4$, $30$, $12$, $10)$ for MRST3, $(\beta$, $\gamma_1$, $\gamma_2$, $\gamma_3$, $\gamma_4$, $\gamma_5)$ $=$ $(4\times10^4$, $30$, $20$, $10$, $10$, $10)$ for MRST5, and $(\beta$, $\gamma_1$, $\gamma_2$, $\gamma_3$, $\gamma_4$, $\gamma_5$, $\gamma_6$, $\gamma_7)$ $=$ $(7\times10^4$, $30$, $12$, $10$, $10$, $10$, $10$, $10)$ for MRST7, respectively.

To compare the performance between various models quantitatively, we compute the root mean square error (RMSE), peak signal to noise ratio (PSNR), structural similarity index measurement (SSIM) of the reconstructions in a region of interest (ROI). The RMSE in Hounsfield units (HU) computed between the ground truth image and reconstruction image is defined as RMSE $= \sqrt{\Sigma_{i \in \text{ROI}}(\hat{x}_i-x^*_i)^2/{N_{\text{ROI}}}}$, where $\hat{x}_i$ and $x^*_i$ denote the pixel intensities of the reconstructed and ground truth images, respectively, and $N_{\text{ROI}}$ is the number of pixels in the ROI. The ROI here was a circular (around center) region containing all the phantom tissues. Since the full-dose Mayo Clinic CT images still contain some considerable noise, we cannot solely rely on RMSE and PSNR as metrics of image quality, as they would include the noise. SSIM helps better evaluate the preservation of structural details in the reconstructed image.

We conduct experiments on three slices (L067-slice100, L192-slice100, L506-slice100) of the Mayo Clinic data. Fig.~\ref{fig:recon_mayo} shows the reconstruction of L067-slice100 using FBP, PWLS-EP, PWLS-ST, PWLS-MRST2, PWLS-MRST3, PWLS-MRST5, and PWLS-MRST7, respectively at incident photon intensity $I_0=10^{4}$. TABLE~\ref{tab:MRST} lists the RMSE, PSNR, and SSIM values of reconstructions of the three test slices, with the best values bolded. The two-layer model (MRST2) provides the best RMSE and PSNR values among the methods. However, when we consider the SSIM criterion, MRST5 and MRST7 outperform ST and MRST2. So which MRST model is better?
 By observing the reconstructed images, we see that although MRST2 and ST have lower RMSE and higher PSNR values than MRST5 and MRST7, they sacrifice some sharpness of the central region and suffer from loss of details. The deeper models have a more positive effect in maintaining subtle features, which is clearly more essential to clinic medical diagnosis. Furthermore, after considerable parameter tuning, we have observed that the deeper models offer more stable image quality as $\beta$ is varied, i.e., they are more robust to oversmoothing.
\begin{table}[!h]	
	\centering
	\caption{RMSE in HU (first row), PSNR in dB (second row) and SSIM (third row) of reconstructions with FBP, PWLS-EP, PWLS-ST, PWLS-MRST2, PWLS-MRST3, PWLS-MRST5, and PWLS-MRST7, for three slices of Mayo Clinic data at incident photon intensity $I_0 = 1\times10^4$.}
	\label{tab:MRST}	 	
	\vspace{-0.05in}
	\renewcommand\tabcolsep{5.0pt}
	\footnotesize{
	\begin{tabular}{c|ccccccc}		
		\toprule
	    	&FBP  & EP &ST  &MRST2    &MRST3   & MRST5  & MRST7  \\
		\midrule
		 \multirow{4}{*}{\shortstack{L067\\slice100}} & 101.1  & 34.2  & 30.3  & \textbf{29.1}  & 29.8   & 30.5  & 31.4 \\
		 \cmidrule{2-8}
		  & 27.4   & 36.8  & 37.8   & \textbf{38.2}  & 38.0   & 37.8  & 37.5 \\
		 \cmidrule{2-8}
		 & 0.359  & 0.728  & 0.724   & 0.731  & 0.731   & \textbf{0.732}  & 0.731\\
		\midrule
		\multirow{4}{*}{\shortstack{L192\\slice100}}  & 60.0  & 26.6  & 23.5   & \textbf{22.2}  & 22.3   & 23.0  & 23.8 \\
		\cmidrule{2-8}
		& 32.4  & 38.9   & 39.9   & \textbf{40.4}  & \textbf{40.4}   & 40.1   & 39.8\\
		\cmidrule{2-8}
		 & 0.440  & 0.786  & 0.790   & 0.798  & 0.800   & 0.807   & \textbf{0.810}\\
		\midrule
		\multirow{4}{*}{\shortstack{L506\\slice100}}  & 56.8  & 32.6  & 27.1   & \textbf{26.0}  & 26.2  & 27.5  & 28.7 \\
		\cmidrule{2-8}
		& 32.8  & 37.6  & 39.2   & \textbf{39.6}  & 39.5  & 39.1  & 38.7 \\
		\cmidrule{2-8}
		 & 0.489  & 0.780  & 0.790   & 0.791  & 0.793   & 0.797  & \textbf{0.802} \\
		\bottomrule
	\end{tabular}
	}
	\vspace{-0.25in}
\end{table}
\vspace{-0.05in}
\section{Conclusion}
This paper proposes a general framework for Multi-layer Residual Sparsifying Transform (MRST) learning wherein the transform domain residual maps over several layers are jointly sparsified. Our work then applies MRST learning to low-dose CT (LDCT) image reconstruction by using a PWLS approach with a learned MRST-based regularizer. Experimental results illustrate the promising performance of the multi-layer scheme over single-layer learned sparsifying transforms. Learned MRST models also offer significant improvement over typical nonadaptive methods. In the future, we will investigate additional strategies such as pooling operations to reduce noise over layers.
\vspace{-0.10in}
\section{Acknowledgment}
The authors thank Dr. Cynthia McCollough, the Mayo Clinic, the American Association of Physicists in Medicine, and the National Institute of Biomedical Imaging and Bioengineering for providing the Mayo Clinic data. 
\vspace{-0.05in}
\bibliographystyle{IEEEbib}	
\footnotesize
\small
\bibliography{refs}	

\begin{thebibliography}{10}

\bibitem{pati:93:omp}
Y.~Pati, R.~Rezaiifar, and P.~Krishnaprasad,
\newblock ``Orthogonal matching pursuit: recursive function approximation with
  applications to wavelet decomposition,''
\newblock in {\em {Asilomar Conf. on Signals, Systems and Computers}}, 1993,
  pp. {40--44 vol.1}.

\bibitem{aharon:06:ksa}
M.~Aharon, M.~Elad, and A.~Bruckstein,
\newblock ``{K-SVD:} an algorithm for designing overcomplete dictionaries for
  sparse representation,''
\newblock {\em IEEE Trans. Sig. Proc.}, vol. 54, no. 11, pp. {4311--4322}, Nov.
  2006.

\bibitem{rubinstein:13:aka}
R.~Rubinstein, T.~Peleg, and M.~Elad,
\newblock ``Analysis {K-SVD}: A dictionary-learning algorithm for the analysis
  sparse model,''
\newblock {\em IEEE Trans. Sig. Proc.}, vol. 61, no. 3, pp. 661--677, Feb.
  2013.

\bibitem{ravishankar:13:lst}
S.~Ravishankar and Y.~Bresler,
\newblock ``Learning sparsifying transforms,''
\newblock {\em IEEE Trans. Sig. Proc.}, vol. 61, no. 5, pp. {1072--1086}, Mar.
  2013.

\bibitem{ravishankar:18:lml}
S.~Ravishankar and B.~Wohlberg,
\newblock ``Learning multi-layer transform models,''
\newblock in {\em {Allerton Conf. on Comm., Control, and Computing}}, 2018, pp.
  {160--165}.

\bibitem{feldkamp:84:pcb}
L.~A. Feldkamp, L.~C. Davis, and J.~W. Kress,
\newblock ``Practical cone beam algorithm,''
\newblock {\em {J. Opt. Soc. Am. A}}, vol. 1, no. 6, pp. {612--619}, June 1984.

\bibitem{elbakri:02:sir}
I.~A. Elbakri and J.~A. Fessler,
\newblock ``Statistical image reconstruction for polyenergetic {X-ray} computed
  tomography,''
\newblock {\em {IEEE Trans. Med. Imag.}}, vol. 21, no. 2, pp. {89--99}, Feb.
  2002.

\bibitem{sauer:93:alu}
K.~Sauer and C.~Bouman,
\newblock ``A local update strategy for iterative reconstruction from
  projections,''
\newblock {\em IEEE Trans. Sig. Proc.}, vol. 41, no. 2, pp. 534--548, Feb.
  1993.

\bibitem{thibault:07:atd}
J-B. Thibault, K.~Sauer, C.~Bouman, and J.~Hsieh,
\newblock ``A three-dimensional statistical approach to improved image quality
  for multi-slice helical {CT},''
\newblock {\em Med. Phys.}, vol. 34, no. 11, pp. {4526--4544}, Nov. 2007.

\bibitem{pfister:14:mbi}
L.~Pfister and Y.~Bresler,
\newblock ``Model-based iterative tomographic reconstruction with adaptive
  sparsifying transforms,''
\newblock in {\em Proc. SPIE}, 2014, vol. 9020, pp. {90200H--1--90200H--11}.

\bibitem{zheng:18:pua}
X.~Zheng, S.~Ravishankar, Y.~Long, and J.~A. Fessler,
\newblock ``{PWLS-ULTRA}: An efficient clustering and learning-based approach
  for low-dose {3D CT} image reconstruction,''
\newblock {\em IEEE Trans. Med. Imag.}, vol. 37, no. 6, pp. 1498--1510, June
  2018.

\bibitem{chun:17:esv}
I.~Y. Chun, X.~Zheng, Y.~Long, and J.~A. Fessler,
\newblock ``Efficient sparse-view {X-ray} {CT} reconstruction using {$\ell_1$}
  regularization with learned sparsifying transform,''
\newblock in {\em {Proc. Intl. Mtg. on Fully 3D Image Recon. in Rad. and Nuc.
  Med}}, June 2017, pp. {115--119}.

\bibitem{ye:19:sld}
S~Ye, S.~Ravishankar, Y.~Long, and J.~A. Fessler,
\newblock ``{SPULTRA}: {Low-Dose CT} image reconstruction with joint
  statistical and learned image models,''
\newblock {\em {IEEE Trans. Med. Imag., (Early Access)}}, Aug. 2019.

\bibitem{zhou:13:atf}
W.~Zhou, J-F. Cai, and H.~Gao,
\newblock ``Adaptive tight frame based medical image reconstruction: a
  proof-of-concept study for computed tomography,''
\newblock {\em {Inverse Prob.}}, vol. 29, no. 12, pp. 125006, Dec. 2013.

\bibitem{nien:16:rla}
H.~Nien and J.~A. Fessler,
\newblock ``Relaxed linearized algorithms for faster {X-ray} {CT} image
  reconstruction,''
\newblock {\em IEEE Trans. Med. Imag.}, vol. 35, no. 4, pp. 1090--1098, Apr.
  2016.

\bibitem{cho:15:rdf}
J.~H. Cho and J.~A. Fessler,
\newblock ``Regularization designs for uniform spatial resolution and noise
  properties in statistical image reconstruction for {3D} {X-ray} {CT},''
\newblock {\em {IEEE Trans. Med. Imag.}}, vol. 34, no. 2, pp. {678--689}, Feb.
  2015.

\end{thebibliography}
\vfill

\end{document}